%% file: main.tex
\documentclass[sigconf,screen,natbib=false]{acmart}

\usepackage{balance}
\usepackage{graphicx}
\AtBeginDocument{%
  }

\copyrightyear{2026}
\acmYear{2026}
\setcopyright{cc}
\setcctype{by}
\acmConference[FSE Companion '26]{34th ACM Joint European Software Engineering Conference and Symposium on the Foundations of Software Engineering}{July 05--09, 2026}{Montreal, QC, Canada}
\acmBooktitle{34th ACM Joint European Software Engineering Conference and Symposium on the Foundations of Software Engineering (FSE Companion '26), July 05--09, 2026, Montreal, QC, Canada}
\acmDOI{10.1145/3803437.3805539}
\acmISBN{979-8-4007-2636-1/2026/07}

\begin{document}

\title{AutonomyLens: A Self-Evolving Simulation-Based Testing Loop for Autonomous Systems}

\author{Ankit Agrawal}
\affiliation{%
  \institution{Department of Computer Science, Saint Louis University}
  \country{USA}
}
\email{ankit.agrawal.1@slu.edu}

\author{Jithin Garapati}
\affiliation{%
  \institution{Department of Computer Science, Saint Louis University}
  \country{USA}
}
\email{jithin.garapati@slu.edu}

\author{Bohan Zhang}
\affiliation{%
  \institution{Department of Computer Science, Saint Louis University}
  \country{USA}
}
\email{bohan.zhang.1@slu.edu}

\begin{abstract}
Software engineering practices for validating autonomous cyber-physical systems (e.g., Uncrewed Aerial Vehicles) remain fragmented across scenario design, simulation execution, and telemetry analysis, limiting traceability between requirements, tests, and evidence. This fragmentation reduces reproducibility, slows debugging and iteration, and hinders systematic assurance under complex and evolving environmental conditions.

We present AutonomyLens, an LLM-driven framework that integrates scenario specification, simulation execution, and telemetry analysis into a unified validation workflow. AutonomyLens enables developers to translate high-level validation intent into executable, temporally evolving scenarios, automatically run simulations, and perform context-aware analysis of resulting system behavior. The framework introduces (i) a structured representation for mission-level scenarios, (ii) an automated execution pipeline, (iii) analysis mechanisms that align telemetry with scenario context to produce actionable insights, and (iv) counterfactual scenario generation that closes the loop by refining and synthesizing new test cases from observed failures. We describe the early-stage design of AutonomyLens, discuss key challenges in building integrated validation workflows for autonomous systems, and outline how such an approach can improve traceability, reproducibility, and scalability in autonomy validation.
\end{abstract}



\maketitle

\input{sections/introduction}
\input{sections/methods}

\input{sections/roadmap}

\balance

\bibliographystyle{ACM-Reference-Format}
\bibliography{software}

\end{document}

%% file: sections/introduction.tex
\section{Introduction}

Software systems controlling autonomous cyber-physical platforms increasingly operate in uncertain physical environments where failures carry safety and societal consequences. Ensuring their reliability requires extensive testing across diverse environmental conditions and corner cases that are difficult, expensive, or unsafe to reproduce in the physical world. While simulation-based validation has become standard practice across domains \cite{zhang2023dronereqvalidator,shah2018airsim,carla}, the software engineering workflows connecting system requirements, simulation scenarios, simulation logs, and data analytics often remain ad hoc and poorly integrated \cite{agrawal2023requirements}.


Model-Based Testing (MBT) \cite{utting2012taxonomy} provides an important foundation for systematic validation by deriving tests from behavioral abstractions and enabling traceability between specifications and execution artifacts \cite{utting2010practical,dalal1999mbt,pretschner2005oneeval}. However, autonomy validation extends beyond traditional MBT assumptions: autonomous systems interact with evolving environments whose dynamics must themselves be modeled and exercised. For instance, validating an autonomous Uncrewed Aerial Vehicle’s (UAV’s) \textit{navigate-to-waypoint} or \textit{return-to-home} logic requires not only modeling discrete mode switches (e.g., takeoff→cruise→approach→land), but also exercising environmental dynamics such as wind gusts, GPS multipath/dropouts, sensor noise, moving obstacles/traffic, and communication delays that directly perturb estimation and control loops. Therefore, system correctness cannot be assessed solely through discrete transitions or pass/fail verdicts, but requires reasoning over continuous behaviors, contextual telemetry, and system-environment interactions. Moreover, testing objectives evolve as observations reshape understanding of system behavior under diverse environmental conditions, calling for feedback-driven test evolution rather than static model exploration.




This paper presents \emph{AutonomyLens}, an autonomy validation framework that generalizes principles from MBT toward environment-aware, feedback-driven validation workflows. A key enabler is recent progress in large language models (LLMs), which can reduce the manual ``glue work'' of validation by helping translate requirements into structured scenario artifacts, interpret heterogeneous execution evidence (e.g., logs, traces, configurations), and synthesize candidate follow-up tests and hypotheses.

This work is at an early architectural stage; we describe the foundations of this paradigm, identify key research challenges, and outline a research agenda toward scalable and developer-centered autonomy validation ecosystems. Section~2 motivates the need for integrated validation tooling, Section~3 presents the AutonomyLens architecture, and Section~4 discusses open concerns and directions for future work.

\section{AutonomyLens Motivation}
Despite a rich ecosystem of simulators across domains (e.g., robotics, autonomous driving, aerial systems), validation of autonomous systems remains challenging at scale \cite{koopman2016challenges}. Scenario-based simulation has significantly advanced validation in domains such as autonomous driving, enabling reproducible evaluation of complex interactions among agents (e.g., vehicles, pedestrians). Platforms such as CARLA \cite{carla} and scenario description languages such as OpenSCENARIO \cite{OpenSCENARIO} and Scenic \cite{fremont2019scenic2} have emerged as de facto standards for structured environments. However, these approaches are largely tailored to systems operating in constrained, rule-governed settings with well-defined semantics (e.g., road networks, traffic rules).

In contrast, many autonomous systems operate in open-ended, unstructured, and dynamic environments—such as aerial, maritime, and mixed-domain settings—where operational conditions differ substantially. These systems must contend with challenges including dynamic environmental disturbances (e.g., wind, currents), degraded or uncertain sensing (e.g., GNSS-denied environments), interactions with heterogeneous agents (e.g., humans, animals, other autonomous systems), and evolving regulatory or mission constraints. Such factors are difficult to represent using existing scenario description languages and are not adequately supported by current simulators.

Moreover, widely used simulation platforms often provide only partial support for validation needs. For instance, some platforms emphasize sensor simulation for perception research but lack mechanisms for modeling temporally evolving environments or for specifying structured, mission-level scenarios. Others require significant manual effort to construct environments, provide limited support for high-fidelity environmental dynamics, and do not support systematic scenario execution or the definition of test oracles. 

As a result, existing tools are insufficient for automated scenario-driven evaluation of autonomy stacks under realistic and evolving mission conditions. We address this gap with AutonomyLens, an LLM-driven framework that enables automated specification, execution, and analysis of temporally evolving, mission-level scenarios, supporting scalable and context-aware validation of autonomous systems.

%% file: sections/methods.tex
\section{AutonomyLens Architectural Perspective}
\begin{figure*}
    \centering
    \includegraphics[width=\linewidth]{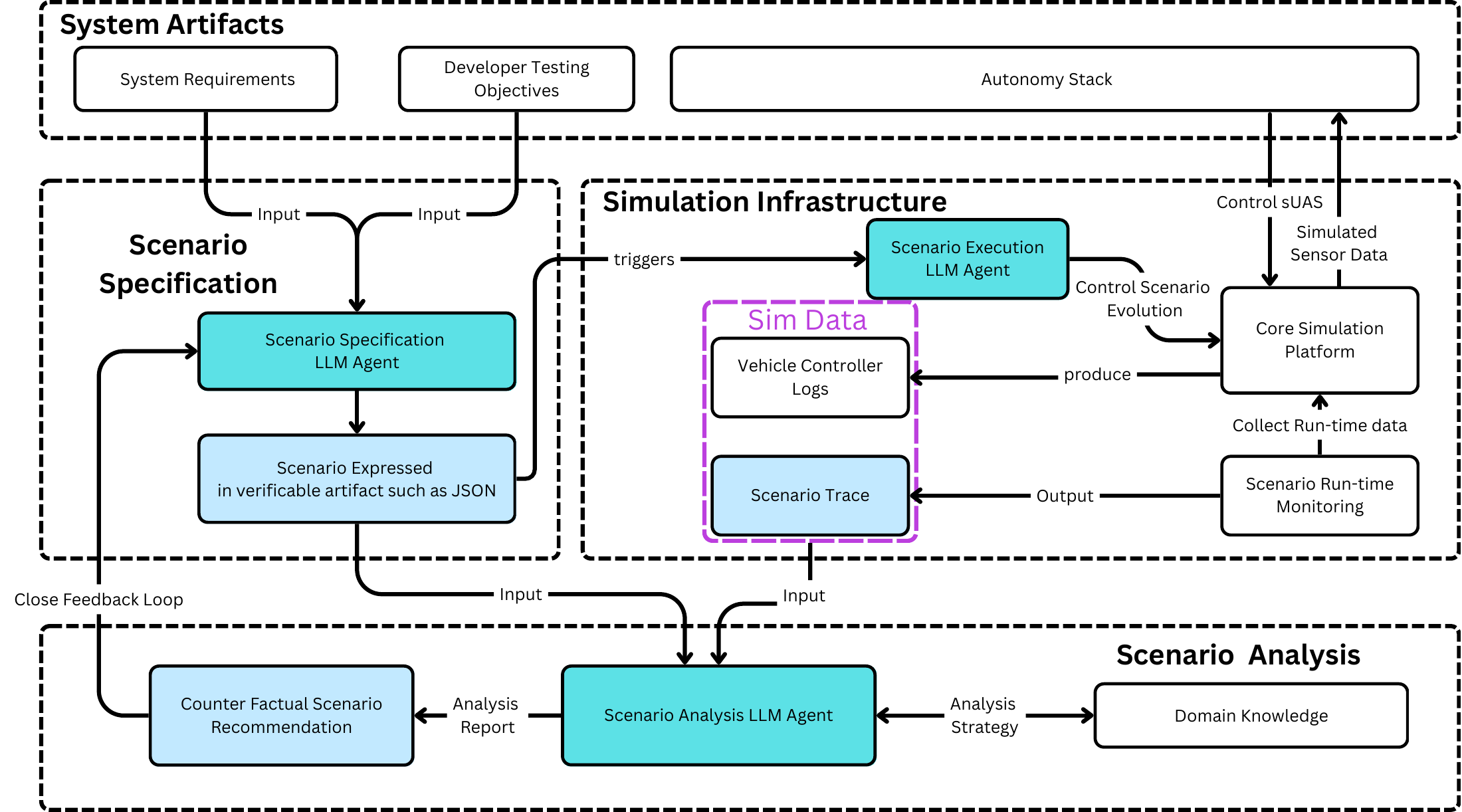}
    \caption{AutonomyLens as a closed-loop validation workflow connecting scenario specification, execution, and analysis through shared artifacts.}
    \Description{Closed-loop architecture diagram showing scenario specification feeding execution, execution producing traces, and analysis generating feedback that refines scenarios.}
    \label{fig:closed_loop_arch}
\end{figure*}
Realizing the vision outlined above requires treating validation not as a sequence of tool invocations but as a structured workflow composed of interacting software artifacts. This artifact-centered design is motivated by our initial interviews with practitioners that automating orchestration and analytics can reduce validation effort by 50\% to 80\%, which requires intent, execution, and evidence to be reusable, machine-actionable artifacts. Therefore, AutonomyLens organizes validation around three artifact transformations—\emph{specification}, \emph{execution}, and \emph{interpretation}, that preserve traceability to developer intent, as well as reproducibility, while enabling iterative refinement driven by observed behavior. Figure~\ref{fig:closed_loop_arch} presents the overall architecture: colored components denote the LLM-driven modules and illustrate how they are embedded within, and interact with, the traditional autonomy validation cycle. A cross-cutting element is \emph{Domain Knowledge}, which supplies platform specifications, environmental models, regulatory constraints, and prior validation experience that inform each stage; for example, wind models and GNSS characteristics guide scenario generation, while flight-controller semantics help the analysis layer interpret telemetry.

\subsection{Scenario Specification}
At the foundation of the workflow is the representation of validation scenarios as structured, versionable artifacts rather than informal scripts. AutonomyLens formalizes scenarios as reviewable objects with explicit semantics that can be validated, transformed, and reused across development cycles.

Therefore, a scenario captures (i) the system objectives to be exercised (e.g., tracking a person in a dense area), (ii) the relevant environmental context (e.g., terrain and built structures, wind fields, and GNSS availability/quality), (iii) dynamic actors in the environment (e.g., other moving people and their characteristics, other aircraft, and ground vehicles), and (iv) stressors that challenge the autonomy stack (e.g., foliage-induced occlusion, gusts, multipath-induced GNSS degradation, rare events, and boundary conditions). Encoding these elements makes scenario intent explicit, inspectable, and machine-actionable.

LLMs can act as an \emph{assistive layer} in this transformation by helping translate requirements and developer intent into structured scenario drafts, suggesting parameterizations and coverage-relevant variations, and documenting assumptions in natural language. Importantly, LLM output is treated as a proposal subject to validation and review, ensuring that the scenario artifact remains explicit and auditable.

\subsection{Simulation Infrastructure}
Developers validate small Uncrewed Aerial System (sUAS) autonomy across simulation tool-chains that differ in world models, physics fidelity, sensor pipelines, and execution semantics \cite{koenig2004design,shah2018airsim}. While most simulators expose the necessary primitives---APIs to spawn actors, apply wind disturbances, perturb sensors (e.g., GNSS dropouts), and record telemetry---these capabilities are realized through simulator-specific representations and scripting interfaces. Consequently, even when the validation intent is stable, it must be re-encoded manually for each simulator.

Therefore, AutonomyLens addresses this gap with an \emph{LLM-based compilation and orchestration agent} that translates a simulator-agnostic scenario artifact into target-specific executable configurations and API-level actions. During execution, the agent coordinates environment instantiation, actor interactions, and environmental perturbations, and it enforces controlled execution (e.g., seeds, schedules) to make outcomes comparable across runs. 

Execution produces standardized traces; the key artifact is the \emph{Scenario Trace}, a structured temporal record that includes time-indexed vehicle telemetry and environment evolution, linking vehicle actions and internal state to environmental context to enable reproducible comparison and post-hoc explanation.

\subsection{Analysis-Driven Scenario Refinement}
The final layer of AutonomyLens acts as \emph{cross-artifact reasoning} driven by LLM over (i) the scenario specification, (ii) the compiled execution configuration, and (iii) the resulting scenario traces and system logs. LLMs can act as an assistive analysis agent by summarizing salient trace segments, aligning observations with scenario elements and configuration settings, and proposing candidate hypotheses and follow-up tests. To preserve developer trust, such outputs should be grounded in explicit trace evidence (e.g., cited time windows, events, and signals) and surfaced as actionable suggestions rather than opaque verdicts.

\emph{Scenario Traces} incorporate both system telemetry and time-indexed environment evolution (e.g., wind, GNSS quality, actor interactions), supporting explanations and follow-up tests that are conditioned on scenario context.

As a result, AutonomyLens grounds failures and near-misses in interpretable context and synthesizes follow-up actions (refined scenarios, regression tests, counterfactual variants). Feedback propagates upstream to refine scenarios, closing the loop and turning outcomes into reusable validation knowledge over time.

\subsection{Example Workflow: UAV Search-and-Rescue Validation}

As a concrete example, consider a UAV search-and-rescue mission in a forested canyon. A developer specifies the goal of tracking a missing person under varying wind and GNSS conditions. AutonomyLens represents this intent as a structured scenario artifact, compiles it into simulator-specific execution steps, and produces a Scenario Trace that links telemetry with environment evolution. If the UAV loses the target near dense foliage during simultaneous wind gusts and GNSS degradation, the analysis layer uses trace-grounded evidence to explain the failure and generate counterfactual follow-up scenarios, such as stronger gusts, different target paths, or earlier sensing degradation. This example illustrates how AutonomyLens supports traceable, iterative, and closed-loop validation.

%% file: sections/roadmap.tex
\section{Discussion and Open Concerns}
AutonomyLens frames autonomy validation as an artifact-centered, feedback-driven workflow: the goal is not a pass/fail label, but an evolving body of \emph{evidence} linking scenario intent, execution traces, and developer-facing claims. Drawing on our experience building simulation-based validation tools for sUAS \cite{zhang2023dronereqvalidator,duvvuru2025llm2,agrawal2023requirements} and on practitioner interviews conducted during that work, we identify four open concerns that must be addressed before the architecture can be realized at scale.

\emph{First}, the meaning of a \emph{scenario trace} should be treated as an engineering contract. Raw logs are rarely self-explanatory: the same telemetry can mean different causes depending on mission phase, controller mode, estimator state, and environment. Trace semantics must therefore bind observations to scenario elements, environmental conditions, and system modes, with schema and provenance that remain interpretable under software evolution. Without this contract, evidence cannot be compared across runs and regressions become hard to attribute.

\emph{Second}, analysis generated by LLMs must be \emph{evidence-grounded} and \emph{calibrated} to earn trust. Employing LLMs in safety-relevant pipelines introduces risks including hallucinated explanations, misinterpretation of scenario semantics, and overconfident diagnoses that may mislead developers \cite{duvvuru2025llm2}. LLMs can reduce triage effort but may rationalize multi-causal failures. To mitigate these risks, explanations should anchor to checkable trace evidence (events, time windows, signals), surface alternative hypotheses when evidence is ambiguous, and state uncertainty explicitly. Suggested follow-up test scenarios should be executable and falsifiable, and all LLM outputs should be treated as proposals subject to developer review rather than authoritative verdicts.

\emph{Third}, deciding \emph{what to run next} in a closed-loop architecture is a resource-allocation problem under finite budgets. Next-test selection should combine environment-centric coverage, utility-aware prioritization that balances discovery vs.\ isolation, and risk-sensitive stopping criteria for when accumulated evidence supports a requirement claim or hazard argument. Otherwise, test evolution devolves into costly iteration without convergence.

\emph{Finally}, evaluation should treat AutonomyLens as a \emph{workflow} rather than a point solution. Beyond generation quality, the key outcomes are reproducibility, time-to-diagnosis, stability of trace semantics under evolution, and the utility of generated follow-up tests. This motivates shared benchmarks pairing scenario intent with expected evidence and diagnostic expectations, alongside longitudinal studies of developer adoption.

In that context, AutonomyLens is a missing software engineering layer for practitioners to reduce week-long setup bottlenecks through automated orchestration and analytics around the simulators teams already use.